\documentclass[aps,groupedaddress,twocolumn,showpacs,prb,floatfix]{revtex4} 
\usepackage{graphicx,rotating,subfigure,amsmath,amsfonts,amssymb,delarray}
\renewcommand{\vec}[1]{\boldsymbol #1}
\newcommand{\e}{\text{e}}
\newcommand{\im}{\text{i}}
\def\l{\left}
\def\r{\right}
\def\12{\frac{1}{2}}
\def\Sr{Sr$_2$CuO$_3\,$}
\def\nn{\nonumber}
\def\eps{\epsilon}

\begin{document}
\bibliographystyle{apsrev}


\title{Spin diffusion and the anisotropic spin-$1/2$ Heisenberg chain}


\author{J. Sirker}
\affiliation{Department of Physics and Astronomy, University of British
  Columbia, Vancouver, British Columbia, Canada V6T 1Z1}

\date{\today}

\begin{abstract}
  Measurements of the spin-lattice relaxation rate $1/T_1$ by nuclear magnetic
  resonance for the one-dimensional Heisenberg antiferromagnet Sr$_2$CuO$_3$
  have provided evidence for a diffusion-like contribution at finite
  temperature and small wave-vector. By analyzing real-time data for the auto-
  and nearest-neighbor spin-spin correlation functions obtained by the
  density-matrix renormalization group I show that such a contribution indeed
  exists for temperatures $T>J$, where $J$ is the coupling constant, but that
  it becomes exponentially suppressed for $T\ll J$. I present evidence that
  the frequency-dependence of $1/T_1$ in the Heisenberg case is smoothly
  connected to that in the free fermion case where the exponential suppression
  of the diffusion-like contribution is easily understood.
\end{abstract}

\maketitle 
\section{Introduction}
\label{intro}
Knowledge about the dynamical properties of the spin-$1/2$ Heisenberg chain
$H=J\sum_j \vec{S}_j\vec{S}_{j+1}$, where $J>0$ is an antiferromagnetic
coupling constant, is the link between theory and many experiments on
compounds which are believed to be good realizations of this model. Whereas
the static properties of the one-dimensional Heisenberg model are well
understood based on effective low-energies theories and its Bethe Ansatz
integrability, many open problems concerning the dynamical properties have
only been addressed very recently. An important question, much interest has
focused on, is how the integrability of the pure model as well as small
integrability-breaking perturbations in any real material influence the spin
(electrical)
\cite{Zotos,RoschAndrei,AlvarezGros,FujimotoKawakami,KluemperJPSJ} and heat
conductivity.\cite{KluemperSakai,JungRosch} A related question important for
nuclear magnetic resonance (NMR), neutron scattering and Coulomb drag between
quantum wires refers to the dynamic spin structure factor. A detailed analysis
of its lineshape at zero temperature $T$ and small wave-vector $q$ has been
presented recently in Refs.~\onlinecite{Glazman,PereiraSirker,Glazmannew}.

Experimentally, most efforts have been concentrated on the compound
Sr$_2$CuO$_3$ which is believed to be an almost ideal realization of a
one-dimensional spin-$1/2$ Heisenberg model with a large in-chain
antiferromagnetic coupling constant $J\sim 2000$ K and very small inter-chain
couplings leading to a N\'eel temperature $T_N\approx 5$ K $\sim 0.003\,J$.
Its Heisenberg character is supported by measurements of the uniform
susceptibility at low temperatures which are compatible with a logarithmic
decrease expected due to marginally irrelevant Umklapp
scattering.\cite{egg94,MotoyamaEisaki,ThurberHunt} Measurements of the thermal
conductivity have revealed strong spatial anisotropies and large parts of the
heat current along the chain direction have been attributed to magnetic
excitations.\cite{SologubenkoGianno} In a pure Heisenberg model the heat
current is a conserved quantity \cite{ZotosPrelovsek} leading to an infinite
thermal conductivity $\kappa_{th}$.\cite{KluemperSakai} The effect of small
integrability-breaking perturbations has been investigated in
Ref.~\onlinecite{JungRosch} and it has been found that $\kappa_{th}$ can
remain anomalously large under certain circumstances. A possible explanation
of the \Sr conductivity data has been proposed in
Ref.~\onlinecite{RozhkovChernyshev} arguing that phonon and impurity mediated
relaxation processes dominate. Another important test of the dynamical
properties of this system at small frequencies $\omega$ has been provided by
NMR measurements of the spin-lattice relaxation rate $1/T_1$. Particularly
appealing is the possibility to separate the contributions from wave-vectors
$q\sim 0$ and $q\sim\pi$, which are the dominant ones for low temperatures, by
measuring $1/T_1$ at inequivalent lattice sites with different form
factors.\cite{TakigawaMotoyama96,ThurberHunt} Theoretical studies of the
spin-lattice relaxation rate have so far been based on the calculation of the
dynamical structure factor in the framework of low-energy effective
theories\cite{Schulz86,SachdevNMR} or the numerical calculation of
imaginary-time correlation
functions.\cite{Sandvik95,StarykhSandvik,TakigawaStarykh,NaefWang} Recent
progress in the calculation of real-time correlation functions by the
density-matrix renormalization group (DMRG) both at zero
temperature\cite{WhiteFeiguin} and finite
temperature\cite{SirkerKluemperDTMRG} has opened a new and so far unexplored
avenue to tackle this problem.

In this article I will focus on the $^{17}$O NMR measurements in
Sr$_2$CuO$_3$\cite{ThurberHunt} where evidence for a $q\sim 0$ mode with
diffusion-like character at finite temperature has been found. To test whether
or not the spin-lattice relaxation rate behaves indeed qualitatively different
in the Heisenberg than in the free fermion case, I will consider the
$XXZ$-model which interpolates between these two cases. In Sec.~\ref{Basics}
the basic theoretical framework to study spin-lattice relaxation will be layed
out
and predictions by the Luttinger model discussed. In Sec.~\ref{fF} the free
fermion case is considered in detail. The interacting case is then analyzed in
Sec.~\ref{numerics} based on real-time data for spin-spin correlation
functions obtained by the DMRG method applied to transfer
matrices.\cite{SirkerKluemperDTMRG} In the last section I discuss and
summarize my main conclusions.

\section{Basic theoretical framework}
\label{Basics}
The Hamiltonian of the $XXZ$-chain is given by
\begin{equation}
H=J\sum_{j=1}^{N}\left[S_{j}^{x}S_{j+1}^{x}+S_{j}^{y}S_{j+1}^{y}+\Delta
  S_{j}^{z}S_{j+1}^{z}-hS^z_j\right]\; ,
\label{XXZ}
\end{equation}
where $J>0$ is an antiferromagnetic coupling constant and $h$ the magnetic
field. $\Delta$ parameterizes an exchange anisotropy and the model is critical
for $-1<\Delta\leq 1$. By Jordan-Wigner transformation the model can be
represented in terms of fermionic operators $\psi_j$
\begin{eqnarray}
\label{XXZ2}
H &=& J\sum_{j=0}^{N}\left[ \frac{1}{2}\left(\psi^\dagger_j \psi_{j+1}+\psi^\dagger_{j+1}\psi_j
    \right)-h\left(\psi^\dagger_j \psi_j
    -\frac{1}{2}\right)\right. \nonumber \\ 
 &+& \left.\Delta \left(\psi^\dagger_j \psi_j
    -\frac{1}{2}\right)\left(\psi^\dagger_{j+1} \psi_{j+1}
    -\frac{1}{2}\right)\right] \; .
\end{eqnarray}
The fermions become non-interacting for $\Delta=0$. We assume for simplicity
that the hyperfine interaction
\begin{equation}
\label{HF}
H_{hf} = \sum_{r} A_{r} \vec{I}_0 \vec{S}_r
\end{equation}
between the nuclear spin $\vec{I}_0$ and the surrounding electron spins
$\vec{S}_r$, where $r$ is the distance in units of the spacing between the
electron spins, is isotropic. If the hyperfine interaction is the dominant
relaxation process, the spin-lattice relaxation rate $1/T_1$ can be obtained
by treating $H_{hf}$ as a perturbation inducing transitions between the
nuclear levels leading to\cite{Moriya56} 
\begin{equation}
\label{T1_1}
\frac{1}{T_1}=\frac{1}{2}\sum_{r,r'}A_rA_{r'}\int_{-\infty}^{\infty}dt\,
\e^{i\omega_N t} \langle S_r^+(t)S_{r'}^-(0)\rangle \; .
\end{equation}
Here $\omega_N$ is the nuclear magnetic resonance frequency with $\omega_N\ll
T$ in all NMR experiments. By Fourier transform we obtain
\begin{equation}
\label{T1_2}
\frac{1}{T_1}=\frac{1}{2}\int\frac{dq}{2\pi} |A(q)|^2 S^{+-}(q,\omega_N) \; ,
\end{equation}
where the transverse dynamic spin structure factor is defined by
\begin{eqnarray}
\label{strucFac}
S^{+-}(q,\omega_N) &=& \sum_j\e^{-\im qj} S^{+-}_j(\omega_N) \\
S^{+-}_j(\omega_N) &=& \int_{-\infty}^\infty dt\,\e^{\im\omega_N t} \langle
S^+_j(t)S^-_0(0)\rangle \nn 
\end{eqnarray}
and 
\begin{equation}
\label{Aq}
A(q)=\sum_r \e^{iqr} A_r \; .
\end{equation}
If the hyperfine interaction (\ref{HF}) is anisotropic, we have to replace
$|A(q)|^2\to (|A_x(q)|^2+|A_y(q)|^2)/2$ in Eq.~(\ref{T1_2}). In spin-chain
compounds there is usually no exchange anisotropy, i.e., $\Delta=1$. In this
case we can replace the transverse by the longitudinal dynamic structure
factor leading to 
\begin{equation}
\label{T1_3}
\frac{1}{T_1}=\int\frac{dq}{2\pi} |A(q)|^2 S^{zz}(q,\omega_N) \; .
\end{equation}
Note, that the spin-spin correlation functions here in principle have to be
evaluated for finite magnetic field. In experiments, however, we often have
the situation that $T\gg |h|$ so that the effect of the magnetic field on the
electron spins can be ignored.

\begin{figure}[!htp]
  \includegraphics[width=0.7\columnwidth]{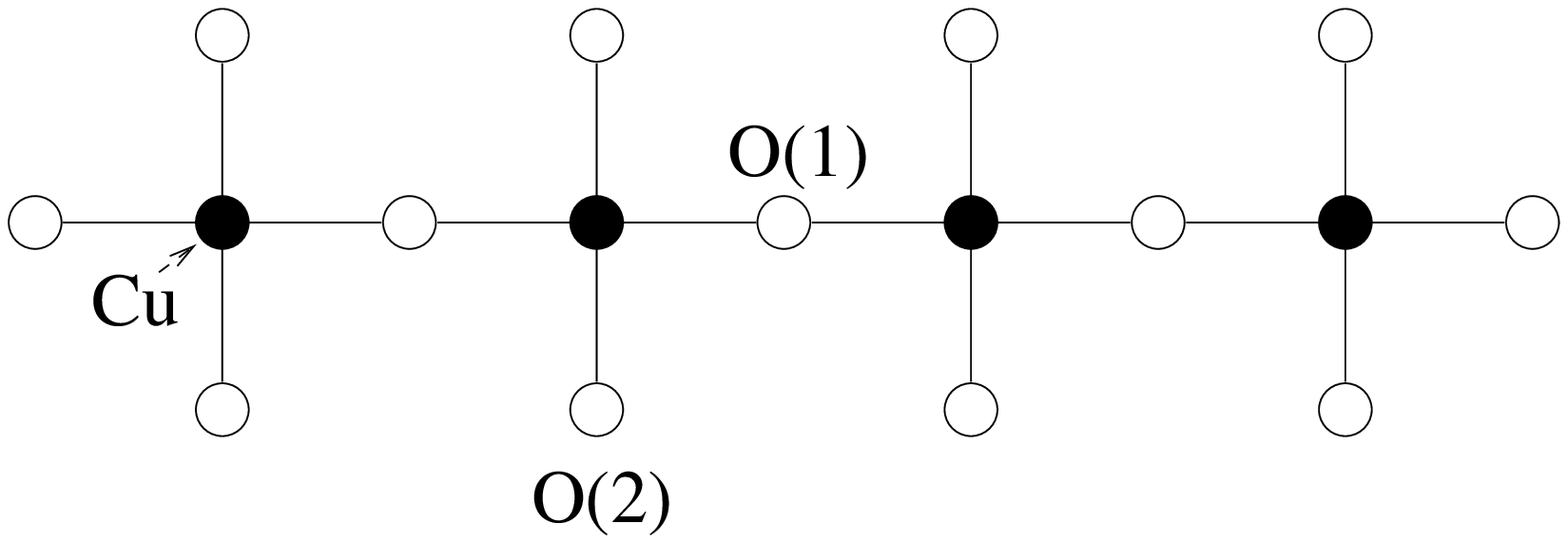}
\caption{Schematic sketch of the spin chain \Sr.}
\label{fig1}
\end{figure}

In \Sr, measurements of the spin-lattice relaxation rate have been performed on
the copper sites,\cite{TakigawaMotoyama96} which carry the $S=1/2$ electron
spin, as well as on the two inequivalent oxygen sites O(1) and
O(2)\cite{ThurberHunt} (see Fig.~\ref{fig1}).
The hyperfine interaction drops down rapidly with distance. It is therefore
reasonable to assume that only the on-site hyperfine interaction $A_0$ and
nearest-neighbor hyperfine interaction $A_1$ are important leading to
$A_{Cu}(q)=\int dq \exp(\im q r) [A_0\delta(r)+A_1\delta(r\pm 1)]= A_0
+ 2A_1\cos(q)$. For the O(2)-site, on the other hand, one finds
\begin{equation}
\label{FormO2}
|A_{O(2)}(q)|^2 = |A|^2
\end{equation}
and for the O(1)-site
\begin{equation}
\label{FormO1}
|A_{O(1)}(q)|^2 = |B|^2 \cos^2(q/2) \; ,
\end{equation}
where $A,B$ are material dependent constants. The spin-lattice relaxation rate
at the O(2)-site can therefore be expressed as
\begin{equation}
\label{O2}
1/T^a_1 = |A|^2 S^{zz}_0(\omega)
\end{equation}
and the one at the O(1)-site as
\begin{equation}
\label{O1}
1/T^b_1 =\frac{1}{2}|B|^2 [S^{zz}_0(\omega)+S^{zz}_1(\omega)] \; .
\end{equation}
Similarly, the spin-lattice relaxation rate for the copper site is given by a
sum of $S^{zz}_0(\omega)$, $S^{zz}_1(\omega)$ and $S^{zz}_2(\omega)$ but with
prefactors which depend on the ratio $A_1/A_0$.\cite{StarykhSandvik} I will
not consider this case here.

The low-energy excitations of the Hamiltonian (\ref{XXZ2}) have either
momentum $q\sim 0$ or momentum $q\sim 2k_F$, with Fermi momentum $k_F=\pi/2$
in the half-filled case considered here. By linearizing the dispersion around
the two Fermi points and expressing the fermionic operators in terms of bosonic
ones, a technique termed bosonization, the $XXZ$-model becomes equivalent (up
to irrelevant operators) to the Luttinger model\cite{GiamarchiBook}
\begin{equation}
H_{LL}=\frac{v}{2}\int dx\,\left[\Pi^2+\left(\partial_x\phi\right)^2\right] \; .
\label{LL}
\end{equation}
Here, $v$ is the spin-wave velocity, $\phi(x)$ a bosonic field and $\Pi(x)$
its conjugated momentum satisfying $[\phi(x),\Pi(y)]=i\delta(x-y)$. For this
free boson model the dynamic structure can be easily calculated and consists
of a uniform ($q\sim 0$) and a staggered ($q\sim\pi$) part where most of the
spectral weight is concentrated.\cite{Schulz86} At the isotropic point,
$\Delta=1$, the staggered part has been shown to lead to
$1/T^a_1\sim\,\ln^{1/2}(T_0/T)$ at low temperatures where $T_0$ is a
scale.\cite{SachdevNMR} The logarithmic temperature dependence is a
consequence of marginally irrelevant Umklapp scattering. The staggered
component should completely dominate the spin-lattice relaxation rates for the
copper- and the O(2)-sites and the obtained
data\cite{TakigawaMotoyama96,ThurberHunt,TakigawaStarykh} indeed show
reasonable agreement with this theoretical prediction.

For the O(1)-site, the form factor (\ref{FormO1}) leads to a strong suppression
of contributions from $q\sim\pi$ and contributions from $q\sim 0$ should
dominate. The uniform part of the dynamic structure factor for the Luttinger
model (\ref{LL}) at $\Delta=1$ is given by
\begin{eqnarray}
\label{deltaFunc}
S^{zz}_u(q,\omega)&=&\frac{|q|}{2(1-\e^{-\omega/T})}\delta(\omega-v|q|) \\
&\stackrel{T\gg\omega}{\approx}&\frac{T}{2\omega}|q|\delta(\omega-v|q|) \nn \; .
\end{eqnarray} 
Here $v=J\pi/2$ is the spin-wave velocity. The spin-lattice relaxation rate
for the uniform part is then
\begin{equation}
\label{O1_LL}
\l(\frac{1}{T^b_1}\r)_u =\frac{2|B|^2T}{\pi^3J^2}\cos^2\l(\frac{\omega}{\pi
  J}\r)\approx\frac{2|B|^2T}{\pi^3 J^2}\l(1-\frac{\omega^2}{\pi^2 J^2}\r) \; . 
\end{equation}
The $\delta$-function peak in the dynamical structure factor (\ref{deltaFunc})
even at finite temperature is a consequence of Lorentz invariance: A single
boson with momentum $|q|$ always carries energy $\omega=v|q|$. This simple
result will be modified by irrelevant operators neglected in (\ref{LL})
corresponding to band curvature terms. The effect of these terms at zero
temperature has been analyzed in
Refs.~\onlinecite{Glazman,PereiraSirker,Glazmannew}. It is, however, not
obvious how to generalize these results to temperatures $T\gg\omega$.

The staggered part of the dynamical structure factor at $\Delta=1$ is given
by\cite{Schulz86}
\begin{equation}
S^{zz}_{s}(q,\omega)=\frac{D}{\omega}\frac{\Gamma\l(\frac{1}{4}-\im\frac{\omega-vq}{4\pi
      T}\r)\Gamma\l(\frac{1}{4}-\im\frac{\omega+vq}{4\pi T}\r)}{\Gamma\l(\frac{3}{4}-\im\frac{\omega-vq}{4\pi
      T}\r)\Gamma\l(\frac{3}{4}-\im\frac{\omega+vq}{4\pi T}\r)} \; ,
\label{stagg}
\end{equation}
where the amplitude $D=(2\pi)^{-3/2}$ has been determined in
Ref.~\onlinecite{Affleck98}. I ignore multiplicative logarithmic corrections
here because they are not important for the temperature range of interest.
Using again the form factor (\ref{FormO1}) we find $(1/T^b_1)_s\approx
0.063\,|B|^2T^2$ for $\omega\to 0$. It has been pointed out in
Ref.~\onlinecite{ThurberHunt} that the value for $1/T_1^bT$ predicted by
(\ref{O1_LL}) agrees with the extrapolation $T\to 0$ of the experimental data
but that the slope of $1/T_1^bT$ in experiment is an order of magnitude larger
than the one obtained from the $q\sim\pi$ contribution in field theory. Most
important, there is no singular frequency dependence within the Luttinger
model which contradicts the behavior of the spin-lattice relaxation rate at
the O(1)-site
\begin{equation}
\frac{1}{T^b_1T}\sim\,\mbox{const}+\frac{T}{\sqrt{\omega_N}} \; .
\label{TH}
\end{equation}
suggested by Thurber {\it et al.}.\cite{ThurberHunt}

In the rest of this paper we want to analyze whether or not the spin-lattice
relaxation rate at the O(1)-site can show such singular frequency dependence
under the assumptions that (a) \Sr is well described by a pure Heisenberg
model with all other degrees of freedom neglected, and (b) that the hyperfine
interaction causes the only important relaxation process. Following
Eq.~(\ref{O1}) the longitudinal auto- and nearest-neighbor correlation
functions will be studied. Although these correlation functions are not
directly related to the relaxation rate for $\Delta\neq 0$, it is helpful to
consider this more general case because it interpolates between the exactly
solvable free fermion and the isotropic Heisenberg model we are interested in.
\section{Free spinless fermions}
\label{fF}
For $\Delta=0$ the Jordan-Wigner transformation yields
\begin{eqnarray}
\label{f1}
&&S^{zz}_r(t)\equiv \langle S^z(r,t)S^z(0,0)\rangle \\
&=& \langle
[c^\dagger(r,t)c(r,t)-1/2][c^\dagger(0,0)c(0,0)-1/2]\rangle \nn \\
&=& \langle c^\dagger(r,t)c(0,0)\rangle\langle c(r,t)c^\dagger(0,0)\rangle \nn
\\
&=& \frac{1}{(2\pi)^2}\int_{-\pi}^\pi \!\!\!\!\! dk_1\int_{-\pi}^\pi \!\!\!\!\! dk_2
\e^{\im(k_1-k_2)r}\e^{\im(\eps_{k_1}-\eps_{k_2})t} n_{k_1} (1-n_{k_2}) \; ,  \nn
\end{eqnarray}
where $\eps(k)= -J\cos k$ and $n_k = (\e^{-J\beta\cos k} +1)^{-1}$. Setting
$J\equiv 1$, the Fourier transform of the autocorrelation function is given
by
\begin{eqnarray}
\label{f2}
&& S^{zz}_0(\omega)  \\
 &=& \frac{1}{2\pi}\int_{-\pi}^\pi \int_{-\pi}^\pi\!\!\!\!\!
dk_1\, dk_2 \,
\delta(\omega-\cos k_1 +\cos k_2) n_{k_1} (1-n_{k_2}) \nn \\
&=& \frac{2}{\pi}\int_{-1+\frac{\omega}{2}}^{1-\frac{\omega}{2}} d\eps
\frac{\theta(2-\omega)}{\sqrt{[1-(\eps-\omega/2)^2][1-(\eps+\omega/2)^2]}} \nn \\
&\times& \frac{1}{\l(\e^{-\beta(\eps+\omega/2)}+1\r)\l(\e^{\beta(\eps-\omega/2)}+1\r)} \nn
\end{eqnarray}
and for the nearest-neighbor correlation function by
\begin{eqnarray}
\label{f3}
S^{zz}_1(\omega) &=&
\frac{2}{\pi}\int_{-1+\frac{\omega}{2}}^{1-\frac{\omega}{2}} \!\!\! d\eps
\frac{\theta(2-\omega)}{\sqrt{[1-(\eps-\omega/2)^2][1-(\eps+\omega/2)^2]}} \nn \\
&\times&
\frac{(\eps-\omega/2)(\eps+\omega/2)}{\l(\e^{-\beta(\eps+\omega/2)}+1\r)\l(\e^{\beta(\eps-\omega/2)}+1\r)} 
\; .
\end{eqnarray}
We are interested in the case $\beta=1/T\gg 1$, $\omega\ll 1$ with $\omega\ll
T$.  Then the most important contributions to the integrals in
(\ref{f2},\ref{f3}) come from $\eps\sim 0$ (Fermi points) and from $\eps\sim
\pm 1$ (top and the bottom of the band) leading to 
\begin{eqnarray}
\label{f4}
S^{zz}_0(\omega) &\sim&
\frac{2}{\pi}T+\frac{2\pi}{3}T^3+\frac{\omega}{\pi}+\frac{\pi}{3}\omega T^2
\nn \\
&+& \frac{2}{\pi}e^{-1/T}(\mbox{const}-\ln\omega)
\end{eqnarray}
and 
\begin{eqnarray}
\label{f5}
S^{zz}_1(\omega) &\sim&
\frac{2\pi}{3}T^3-\frac{1}{2\pi}\omega^2 T -\frac{1}{4\pi}\omega^3 +\frac{\pi}{3}\omega T^3
\nn \\
&+& \frac{2}{\pi}e^{-1/T}(\mbox{const}-\ln\omega) \; .
\end{eqnarray}
Here the first lines in (\ref{f4},\ref{f5}) correspond to contributions from
the Fermi points and the second lines to the ones from the top and bottom of
the band. In particular, we see that the only term divergent for $\omega\to 0$
becomes exponentially suppressed for temperatures $T<J$. We expect that this
picture will remain valid even in the interacting case but with possibly
renormalized $T$- and $\omega$-exponents. This conjecture will be tested
numerically in the next section. For infinite temperature the integrals
(\ref{f2},\ref{f3}) can be expressed as
\begin{equation}
\label{f6}
S^{zz}_0(\omega)=\frac{K(1-4/\omega^2)}{\pi \omega}
\end{equation}
and
\begin{equation}
\label{f7}
S^{zz}_1(\omega)=\frac{K(1-4/\omega^2)}{\pi\omega}-\frac{\omega}{2\pi}E(1-4/\omega^2) 
\end{equation}
where $K$, $E$ are the elliptic integrals of the first and second
kind, respectively. This confirms the logarithmic frequency dependence for
$\omega\to 0$ in this case. 

Our analysis of the interacting case will be based on real-time numerical data
for $S^{zz}_{0,1}(t)$. In order to calculate the Fourier transform we need to
extrapolate in time. As a guide we will use the long-time asymptotic in the
free fermion case. We can write (\ref{f1}) as
 \begin{eqnarray}
\label{f8}
&&S^{zz}_r(t) \\
&=& \frac{1}{4}\left[\frac{1}{2\pi}\int_{-\pi}^\pi dk \;\e^{\im(kr+\eps_k
    t)}\left(1-\tanh(\beta\eps_k/2)\right)\right] \nn \\
&\times& \left[\frac{1}{2\pi}\int_{-\pi}^\pi dk \;\e^{-\im(kr+\eps_k
    t)}\left(1+\tanh(\beta\eps_k/2)\right)\right] \; .\nn
\end{eqnarray}
For zero and infinite temperature this leads to\cite{Niemeijer,KatsuraHoriguchi} 
\begin{equation}
\label{f9}
S^{zz}_0(t)=\l\{\begin{array}{ll}  \frac{1}{4}\l[J_0(t)-\im H_0(t)\r]^2\;
  ,& T=0 \\[0.2cm]
\frac{1}{4}J_0^2(t)\; , & T=\infty
\end{array} \r.
\end{equation}
and 
\begin{equation}
\label{f10}
S^{zz}_1(\omega)=\l\{\begin{array}{ll}  \frac{1}{4}\l[J_1(t)+\im H_{-1}(t)\r]^2\;
  ,& T=0 \\[0.2cm]
\frac{1}{4}J_1^2(t)\; , & T=\infty
\end{array} \r.
\end{equation}
where $J_n$, $H_n$ are the $n$-th order Bessel and Struve functions,
respectively. The long-time asymptotics of these functions is given by
$J_n(t)\sim \sqrt{2/(\pi t)}\cos(t-n\pi/2-\pi/4)$, $H_0(t)\sim \sqrt{2/(\pi
  t)}\sin(t-\pi/4)+2/(\pi t)$ and $H_{-1}(t)\sim \sqrt{2/(\pi
  t)}\sin(t+\pi/4)$. It is instructive to derive the long-time asymptotics for
(\ref{f8}) directly in the $T=0$ case. For this purpose, consider the integral 
\begin{equation}
\label{f11}
I(r,t)=\frac{1}{\pi}\int_{-\pi/2}^{\pi/2} dk \;\e^{\im(kr-t\cos k)} \; .
\end{equation}
For $t>r$, the integral has a saddle point at $k=-\arcsin(r/t)$. That means
that $I(r,t)$ for $t\to\infty$ is dominated by contributions from $k\sim 0$.
These contributions can be evaluated by steepest descend methods. For $t<r$
and $r\gg 1$ the most important contributions come from the Fermi points
$k\sim \pm \pi/2$. Taking both contributions into account leads to
 \begin{eqnarray}
\label{f12}
I(r,t)&\sim& \sqrt{\frac{2}{\pi t}}\exp[-\im(t-\pi/4)]\exp[-\im r^2/(2t)] \\
&+& \frac{\im}{\pi}\frac{\exp(-\im\pi
  r/2)}{t+r}+\frac{\im}{\pi}\frac{\exp(\im\pi r/2)}{t-r} \; , \nn
\end{eqnarray}
where the first line is the $k\sim 0$ contribution and the second line the one
from $k\sim \pm \pi/2$. For the second integral in (\ref{f8}) we can do an
analogous calculation leading to 
\begin{eqnarray}
\label{f13}
S^{zz}_r(t)&\sim& \frac{1}{2\pi
  t}\e^{-2\im(t-\pi/4-r\pi/2)}\e^{-\im
  r^2/t} \\
&-& \frac{1}{2\pi^2}\frac{r^2+t^2}{(r^2-t^2)^2}-\frac{(-1)^r}{2\pi^2}\frac{1}{t^2-r^2} \; . \nn
\end{eqnarray}
This result cast some doubt on the field theory results for $1/T_1$ discussed
in the previous section: The second and third term in (\ref{f13}) can also be
obtained by bosonization. In the interacting case, the exponent of the
alternating term as well as the amplitudes of both terms will then become
$\Delta$-dependent. The first term, however, which completely dominates for
large $t$, cannot be obtained by these methods. In other words, field theory
only describes the time-dependence of correlation functions for $r\gg 1$ and
$t<r$.

By a similar calculation we can also obtain the long-time asymptotics at
finite temperature, in particular
\begin{equation}
\label{f14}
S^{zz}_0(t)\sim \frac{1}{2\pi
  t}\left[\cos\l(t-\frac{\pi}{4}\r)-\im\tanh\frac{1}{2T}\sin\l(t-\frac{\pi}{4}\r)\right]^2 
\end{equation}
and
\begin{equation}
\label{f15}
S^{zz}_1(t)\sim \frac{1}{2\pi
  t}\left[\cos\l(t-\frac{3\pi}{4}\r)+\im\tanh\frac{1}{2T}\sin\l(t+\frac{\pi}{4}\r)\right]^2 
  \; .
\end{equation}
 
\section{Numerics}
\label{numerics}
Recent progress has made it possible to study the real-time dynamics in
one-dimensional quantum systems by DMRG methods. Here we want to use a new
variant of the DMRG applied to transfer matrices
(TMRG)\cite{SirkerKluemperDTMRG} to calculate the longitudinal, real-time
auto- and nearest-neighbor correlation functions for the $XXZ$-model. The main
advantage of this method compared to exact
diagonalization\cite{FabriciusMcCoy} is that the thermodynamic limit can be
performed exactly. Therefore time-dependent correlation functions can be
calculated for arbitrary distances over a wide temperature range. The time
range, however, is limited by the fact that the spectrum of the reduced
density matrix used to truncate the Hilbert space becomes dense. In all
calculations presented here $400$ states were kept in the real-time TMRG
algorithm. A detailed analysis of the accuracy of this method and the maximum
times currently achievable has been presented in
Ref.~\onlinecite{SirkerKluemperDTMRG}. In some sense the real-time TMRG method
is complementary to the calculation of imaginary-time correlation functions by
standard Quantum Monte-Carlo (QMC)\cite{Sandvik95,StarykhSandvik} or
TMRG\cite{NaefWang} methods: Imaginary-time correlation functions can be
calculated completely due to periodicity. The difficulties in this approach
arise from the analytical continuation which is an ill-posed problem. In the
real-time domain, on the other hand, the Fourier transform is well defined but
we have to deal with incomplete numerical data.

\begin{figure} 
\begin{center}
\includegraphics*[width=0.99\columnwidth]{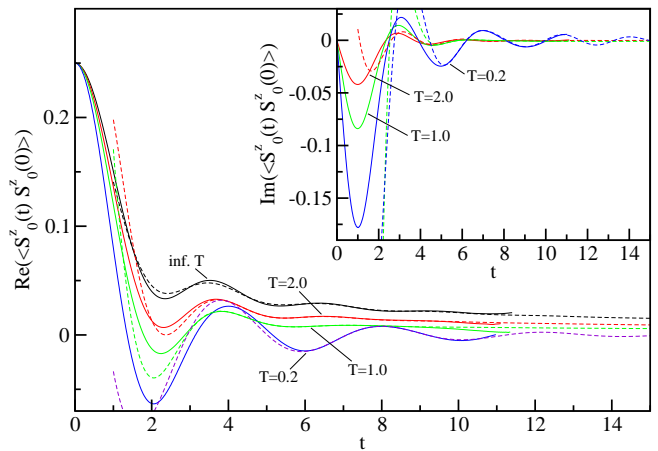} 
\end{center}
\caption{Numerical data (solid lines) for the autocorrelation correlation function at $\Delta=1$ and fits
  (dashed lines) according to Eqs.~(\ref{n1},\ref{n2}).}
\label{fig2}
\end{figure}

To extrapolate in time we will fit the real part of the numerical data for the
auto- and nearest-neighbor correlation function by
\begin{equation}
\label{n1}
f_R(t)=\l[A+B\e^{-\gamma t}\cos(\Omega(t-t_0))\r]/t^d
\end{equation} 
and the imaginary part by
\begin{equation}
\label{n2}
f_I(t)=\l[\tilde{A}\cos(\tilde{\Omega}(t-\tilde{t}_{0})\e^{-\tilde{\gamma}
  t}+\frac{\tilde{B}}{\sqrt{t}}\cos(\tilde{\Omega}_2(t-\tilde{t}_{02}))\r]/t^d \; ,
\end{equation} 
with fit parameters $A,B,\gamma,\Omega,t_0,d$ for the real and
$\tilde{A},\tilde{B},\tilde{\gamma},\tilde{\Omega},\tilde{\Omega}_2,\tilde{t}_0,\tilde{t}_{02},d$
for the imaginary part, respectively. These fit functions are motivated by the
long-time asymptotics in the free fermion case (\ref{f14},\ref{f15}). Note,
that $\gamma=\tilde{\gamma}=0$ in the free fermion case and that next-leading
corrections have been taken into account for the imaginary part. The idea is
to start for each temperature considered with the exactly known fit parameters
in the free fermion case and then increase the interaction $\Delta$ in small
steps thus guaranteeing good start values for the fit parameters for each
anisotropy. As example, the autocorrelation function at $\Delta=1$ for
different temperatures and the corresponding fits are shown in
Fig.~\ref{fig2}.

We begin our detailed analysis with the case $T=\infty$ where the correlation
functions are real. From the results in the previous section for the free
fermion point, we expect that any singular frequency-dependence will be most
pronounced in this limit. Particularly interesting in this context is how the
fit parameter $d$ in (\ref{n1}) evolves as a function of anisotropy $\Delta$
(see Table \ref{tab1}).
 \begin{table}
 \caption{
  The fit parameter $d$ as a function of anisotropy
   $\Delta$ at $T=\infty$ for the auto- and nearest-neighbor
   correlation function, respectively. A variation of the fit
   region gives an error estimate $\sim \pm 0.1$ in all cases. For comparison,
   the values $d^{ED}_{auto}$ obtained in
   Ref.~\onlinecite{FabriciusMcCoy} by a fit to exact diagonalization data for
   a $N=16$ site system are shown.}  
 \begin{ruledtabular}
 \begin{tabular}{cccc}
 $\Delta$ & $d_{auto}$ & $d^{ED}_{auto}\; (N=16)$ & $d_{NN}$\\
 \hline
 0.0 & 1.0 & 1.0     & 1.0 \\
 0.2 & 0.883 & 0.875 & 0.892 \\
 0.4 & 0.774 & 0.835 & 0.769 \\
 0.6 & 0.786 & 0.941 & 0.813 \\
 0.8 & 0.775 & 0.840 & 0.728 \\
 1.0 & 0.683 & 0.705 & 0.643 \\
 \end{tabular}
 \end{ruledtabular}
 \label{tab1} 
 \end{table}
 The numbers obtained clearly show that $d_{auto}$ for the auto- as well as
 $d_{NN}$ for the nearest-neighbor correlation function decrease with
 increasing $\Delta$. Furthermore, $d_{auto}\sim d_{NN}$ for all anisotropies.
 This agrees with the expectation that the power-law decay for $t\gg r$ should
 not depend on the spatial distance $r$. The autocorrelation function in the
 case $T=\infty$ has been investigated previously on the basis of exact
 diagonalization data for chains up to $N=16$ sites.\cite{FabriciusMcCoy}
 There, the same fit function (\ref{n1}) has been used to analyze the long-time
 asymptotics and the exponents obtained show the same trend as a function of
 anisotropy (see Table \ref{tab1}).

\begin{figure}
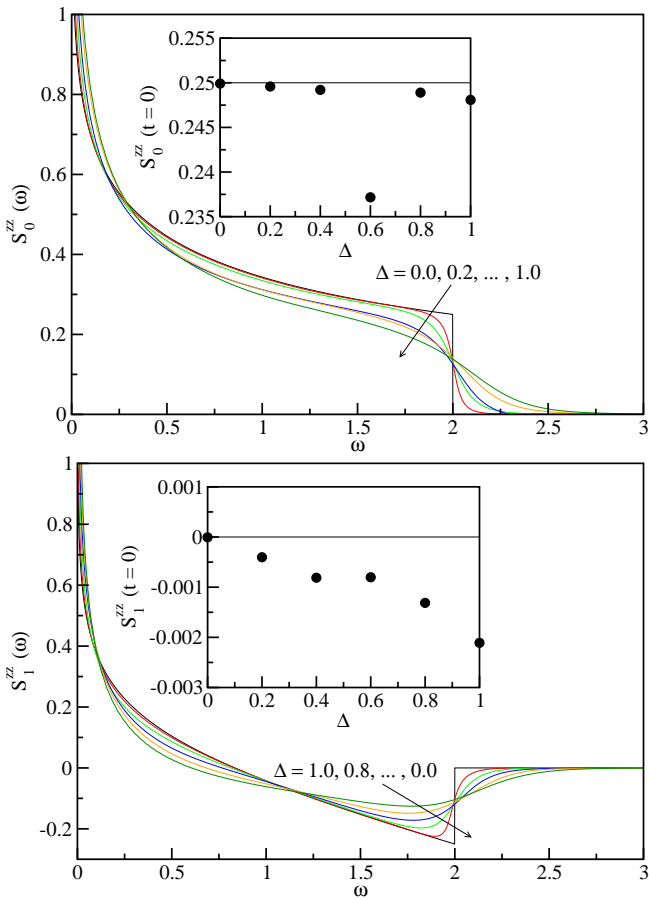
 
\begin{center}
\includegraphics*[width=0.99\columnwidth]{Auto_infT.eps} 
\includegraphics*[width=0.99\columnwidth]{NN_infT.eps} 
\caption{\label{fig3} $S^{zz}_0(\omega)$ and $S^{zz}_1(\omega)$ for infinite
  temperature and different anisotropies. For $\Delta=0$ the exact results
  (\ref{f6},\ref{f7}) are shown. Insets: Test of the sum rule (\ref{Sum}) for
  each anisotropy. The lines denote the exact results, the dots the integrals
  of the numerical data.}
\end{center}
\end{figure}

The extrapolated numerical data can then be Fourier transformed.
The results are shown in Fig.~\ref{fig3}.
At small frequencies both correlation functions show a power-law divergence
$S^{zz}_{0,1}(\omega)\sim \omega^{-\alpha}$ with an exponent $\alpha = 1-d$
and $d$ as in Table \ref{tab1}. In particular, $\alpha\sim 0.3-0.4$ for
$\Delta=1$. Although this does not agree with the phenomenological theory of
spin diffusion by Bloembergen\cite{Bloembergen} and de Gennes\cite{deGennes}
which would predict $\alpha=1/2$, it is already extremely difficult in actual
NMR measurements to determine whether or not the frequency dependence is
singular let alone to determine the exponent. From this perspective, we might
call any kind of divergence a diffusion-like behavior. Using this terminology,
we conclude that there is indeed a diffusion-like contribution to the
spin-lattice relaxation rate at infinite temperature. Another point worth
mentioning is the high-frequency tail in Fig.~\ref{fig3} for $\Delta\neq 0$.
In the free fermion case all excitations contributing to the dynamical
structure factor and therefore to $S^{zz}_{0,1}(\omega)$ are single
particle-hole excitations. The energy of these excitations is limited by the
bandwidth. For the interacting case, however, excitations of multi
particle-hole type are possible which can carry arbitrarily large energies.

A test if the extrapolated real-time data indeed yield reasonable results for
$S^{zz}_{0,1}(\omega)$ is provided by the sum rule
\begin{equation}
\label{Sum}
\int_{-\infty}^\infty \frac{d\omega}{2\pi} S^{zz}_r(\omega) = S^{zz}_r(t=0) \; .
\end{equation}
Because $S^{zz}_r(-\omega) = \exp(-\beta\omega)S^{zz}_r(\omega)$ it is
sufficient to consider the correlation functions for positive frequencies
only. For the autocorrelation function $S^{zz}_0(t=0) = 0.25$ for all
anisotropies and temperatures. For $S^{zz}_1(\omega)$ and finite temperatures,
the integrated intensity has to be compared with numerical data for the static
correlation function. For infinite temperature, however, $S^{zz}_1(t=0) = 0$.
The results of this test are shown in the insets of Fig.~\ref{fig3}.

\begin{figure*}
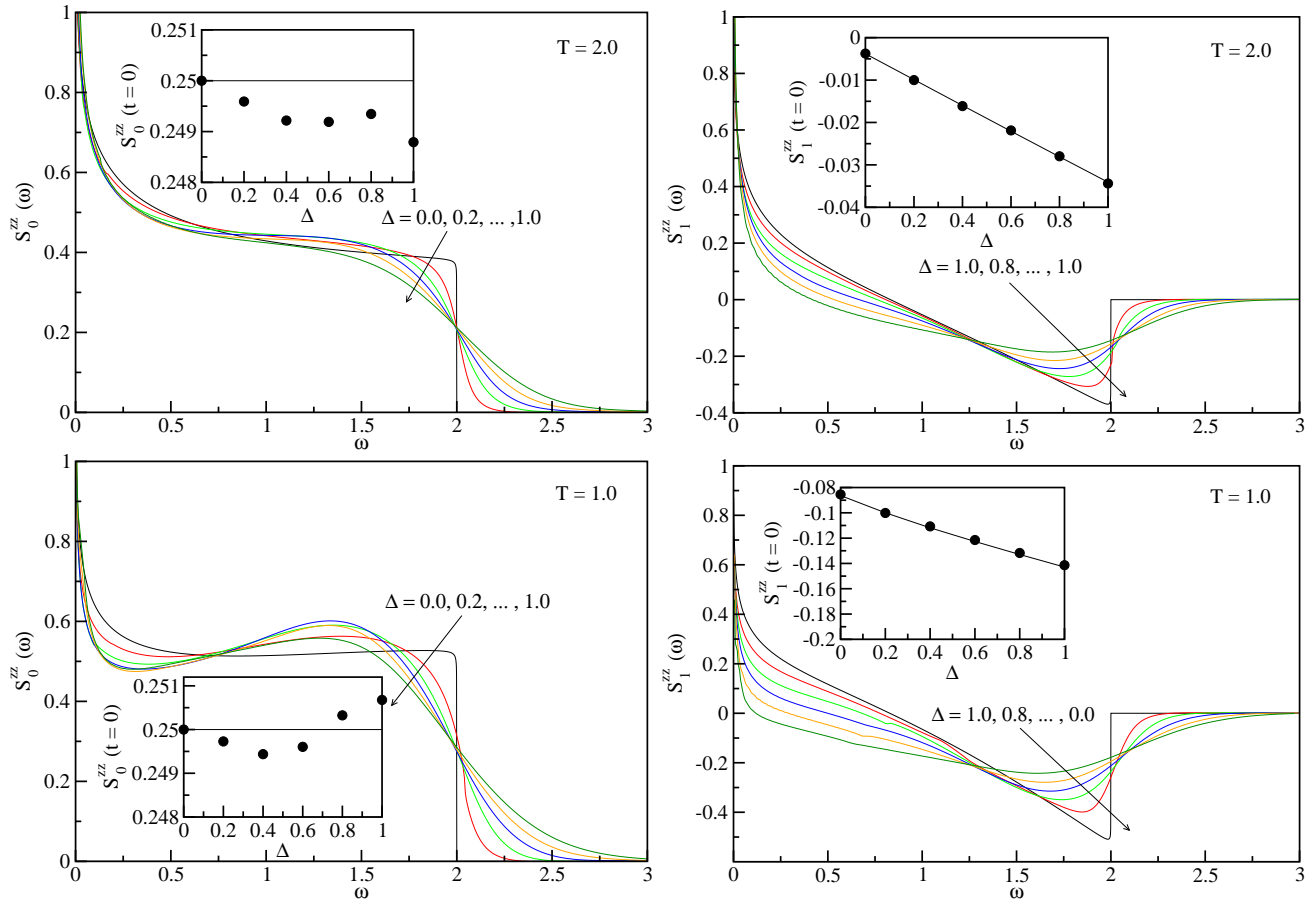
 
\begin{center}
\includegraphics*[width=0.99\columnwidth]{Auto_T2.0.eps} 
\includegraphics*[width=0.99\columnwidth]{NN_T2.0.eps} 
\includegraphics*[width=0.99\columnwidth]{Auto_T1.0.eps} 
\includegraphics*[width=0.99\columnwidth]{NN_T1.0.eps}
\caption{\label{fig4} $S^{zz}_0(\omega)$ and $S^{zz}_1(\omega)$ for $T=2.0,\,
  1.0$ and different anisotropies. For $\Delta=0$ the exact results
    (\ref{f2},\ref{f3}) are shown. In the insets the integrated
    intensities (dots) are compared with the exact result $S^{zz}_0(t=0)=0.25$
    for the autocorrelation and with numerical data for the static
    nearest-neighbor correlation function (lines). }  
\end{center}
\end{figure*}

Next, we consider finite temperatures. Based on the exact solution in the free
fermion case and on what we know from field theory about the contributions from the Fermi
points, it is reasonable to assume that the exponents $d$ in
(\ref{n1},\ref{n2}) are identical and that this exponent depends on anisotropy
only and not on temperature. We have therefore fixed $d=(d_{auto}+d_{NN})/2$
for each anisotropy with $d_{auto},\, d_{NN}$ as in Table \ref{tab1}.\footnote{I
  tested that the curves for $S^{zz}_{0,1}(\omega)$ do not significantly
  change if $d$ is treated as a free parameter.} The Fourier transform of the
extrapolated data for $T=2.0$ and $T=1.0$ is shown in Fig.~\ref{fig4} and a
check of the sum rule (\ref{Sum}) in the insets.
The results are qualitatively similar to the infinite temperature case. In
particular, the same kind of power-law divergencies for $\omega\to 0$ and
$\Delta\neq 0$ are present. Furthermore, a peak in $S^{zz}_0(\omega)$ starts to
develop around $\omega\sim \pi/2$.

\begin{figure*}
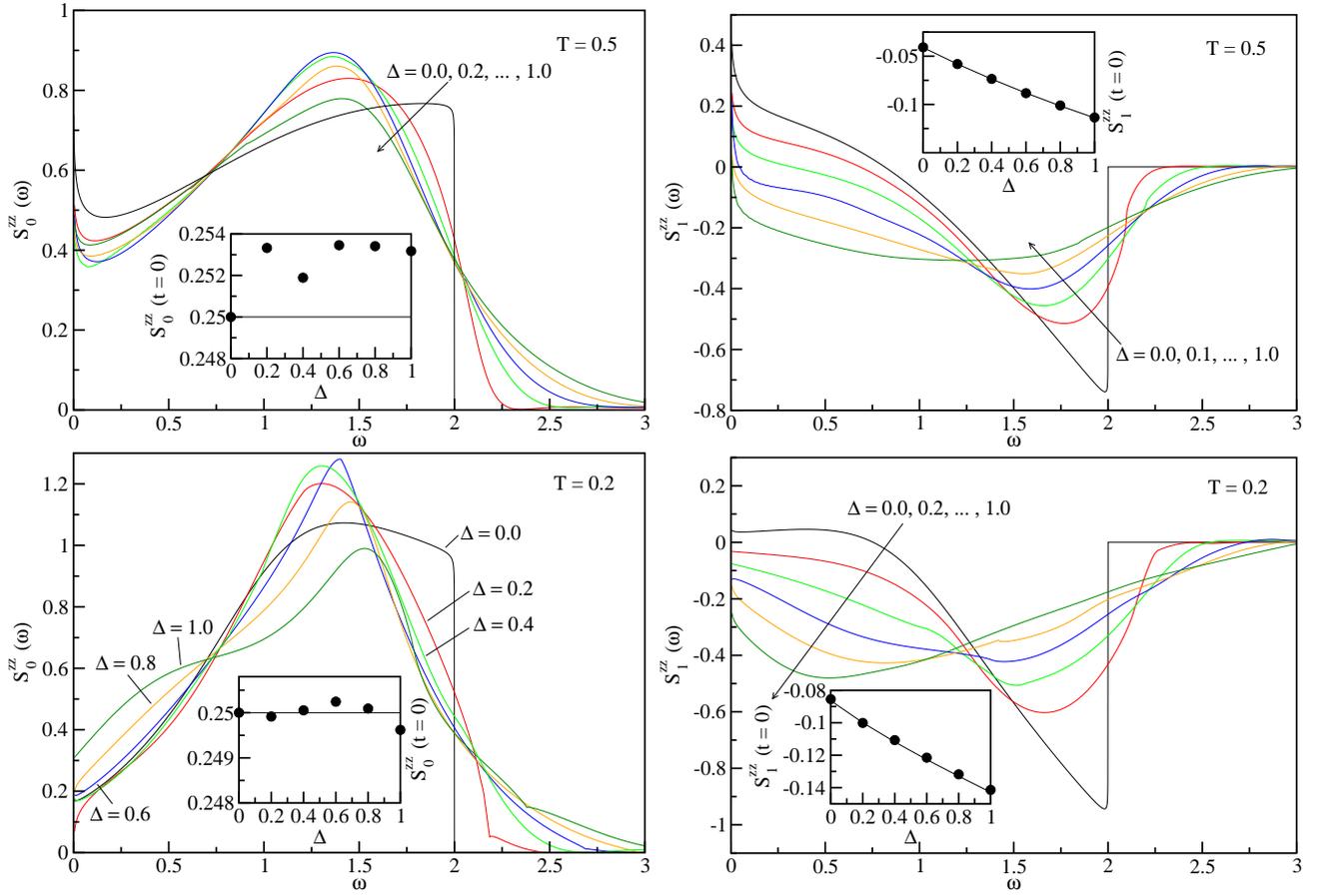
 
\begin{center}
\includegraphics*[width=0.99\columnwidth]{Auto_T0.5.eps} 
\includegraphics*[width=0.99\columnwidth]{NN_T0.5.eps}
\includegraphics*[width=0.99\columnwidth]{Auto_T0.2.eps} 
\includegraphics*[width=0.99\columnwidth]{NN_T0.2.eps}
\caption{\label{fig4b} $S^{zz}_0(\omega)$ and $S^{zz}_1(\omega)$ for $T=0.5,\,
  0.2$ and different anisotropies. For $\Delta=0$ the exact results
    (\ref{f2},\ref{f3}) are shown. The insets show a test of the sum rule
    (\ref{Sum}).}  
\end{center}
\end{figure*}

\begin{figure} 
\begin{center}
\includegraphics*[width=0.99\columnwidth]{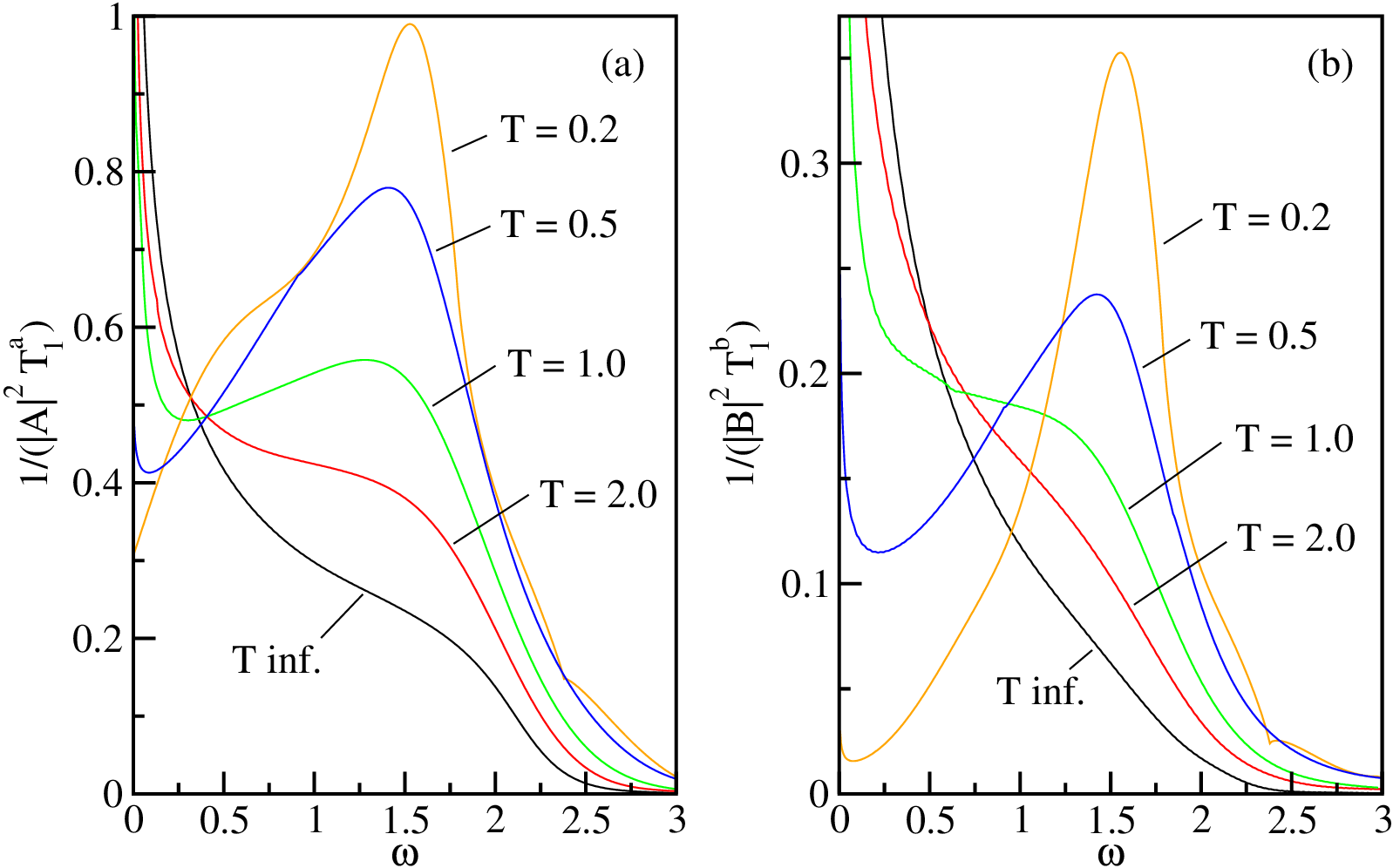} 
\caption{\label{fig5} (a) $1/T^a_1$ for the O(2)-site and (b) $1/T^b_1$ for the O(1)-site in \Sr.}  
\end{center}
\end{figure}

For $T<1$, however, the divergence at small frequencies gets strongly
suppressed in both correlation functions (see Fig.~\ref{fig4b}) and at $T=0.2$
the singular frequency dependence can no longer be detected.  
The small kinks visible in Fig.~\ref{fig4b} for $T=0.2$ are not of physical
origin. They are most likely connected to oscillations in the real-time
numerical data leading to peaks or dips in $S^{zz}_{0,1}(\omega)$ at the
corresponding frequencies. $S^{zz}_0(\omega)$ has been studied previously for
$\Delta=1$ by high-temperature series expansions and QMC\cite{StarykhSandvik}
as well as by a calculation of the imaginary-time correlation function using
the TMRG algorithm.\cite{NaefWang} The results for this special case (see also
Fig.~\ref{fig5}(a)) presented here are very similar to the ones in these
works. For example, we also see a peak at $\omega\sim \pi/2$ which increases
in height with decreasing temperature and the development of a shoulder at
$\omega\sim 0.7$ for $T=0.2$. Quantitatively, however, the data in
Refs.~\onlinecite{StarykhSandvik,NaefWang} are about a factor $2$ smaller for
all frequencies. As the results here fulfill the sum rule (\ref{Sum}) for all
temperatures with good accuracy this suggests that a factor $2$ might be
missing for the numerical data shown in
Refs.~\onlinecite{StarykhSandvik,NaefWang}.

The spin-lattice relaxation rates $1/T^a_1$ and $1/T^b_1$ for $\Delta=1$ can
then be obtained using (\ref{O2},\ref{O1}) and are shown in Fig.~\ref{fig5}.   
The behavior at small frequencies seems to be similar in both cases: There is
a power law divergence $\omega^{-\alpha}$ with $\alpha\sim 0.3-0.4$ at
temperatures $T>J$ (remember that we set $J=1$ here), however, this divergence
gets strongly suppressed at temperatures $T<J$. This suggests that the
contributions to the spin-lattice relaxation rate with singular frequency
dependence behave indeed similar to (\ref{f4},\ref{f5}) found in the free
fermion case.

\begin{figure} 
\begin{center}
\includegraphics*[width=0.99\columnwidth]{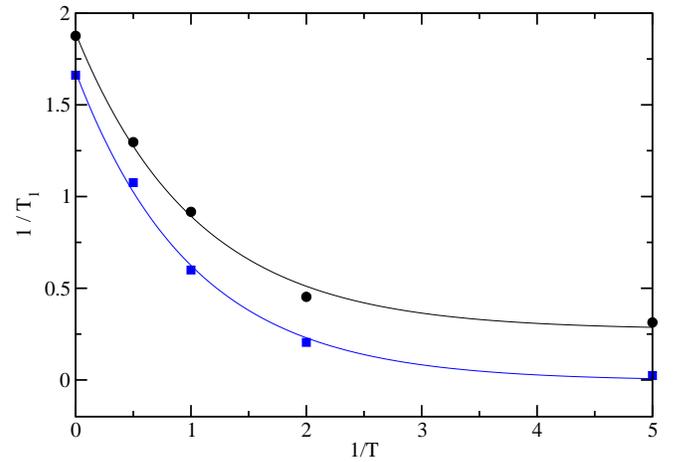} 
\caption{\label{fig6} Temperature dependence of $1/T_1^a$ (black dots) and $1/T_1^b$ (blue
  squares) for $\omega=0.01$ with $A=B=1$ in Eqs.~(\ref{O1},\ref{O2}). The
  black line is a fit $1/T_1^a=0.27+1.62\exp(-0.96/T)$ and the blue line a fit
  $1/T_1^b=0.0+1.69\exp(-0.99/T)$.}
\end{center}
\end{figure}

To analyze the temperature dependence in more detail, $1/T_1$ for a fixed
frequency $\omega\ll T$ is shown in Fig.~\ref{fig6}.
The fits in Fig.~\ref{fig6} show that the spin-lattice relaxation rate
decreases exponentially with temperature in both cases and that the scale of
the exponential decrease is set by $J$ as in the free fermion case. We cannot
analyze the behavior of $1/T_1$ at low temperatures in detail due to
insufficient numerical data. However, the value $1/T_1^a(T=0.2) \approx
0.31|A|^2$ is close to the one predicted by the field theory formula
(\ref{stagg}) if we include logarithmic corrections to
scaling\cite{StarykhSingh} yielding $1/T_1^a(T=0.2) \approx 0.27|A|^2$. We
also note that $1/T_1^a$ only increases by about 30\% when changing the
temperature from $T=0.2$ to $T=0.5$. This is an indication that $1/T_1^a$ will
indeed be almost constant at low temperatures as predicted by field theory.
For the O(1)-site, our numerical data are consistent with $1/T_1^b\to 0$ for
$T\to 0$.

\section{Conclusions}
\label{con}
The purpose of this article has been to investigate if a one-dimensional
Heisenberg antiferromagnet has a diffusion-like contribution to the
spin-lattice relaxation rate as has been proposed in
Ref.~\onlinecite{ThurberHunt} based on $^{17}$O NMR experiments on \Sr. To
tackle this problem I found it useful to consider the more general $XXZ$-case
which interpolates between the exactly solvable free fermion and the
Heisenberg model we are interested in. For the free fermion model I have shown
that a contribution to the spin-lattice relaxation rate exists which diverges
logarithmically for frequency $\omega\to 0$. However, this contribution comes
from the top and the bottom of the band and becomes therefore exponentially
suppressed at temperatures $T<J$. The contributions from the Fermi points, on
the other hand, do not show any singular frequency dependence. I then analyzed
the interacting case based on real-time numerical data for the auto- and
nearest-neighbor correlation functions obtained by the density-matrix
renormalization group applied to transfer matrices.\cite{SirkerKluemperDTMRG}
The advantage of working in the real-time domain compared to imaginary-time
methods is that the ill-defined analytical continuation of numerical data is
circumvented. On the flip side, there is no periodicity in real-time so the
numerical data have to be extrapolated in time before they can be Fourier
transformed. I showed that one can do such an extrapolation using the
long-time asymptotics in the free fermion case as a guide. I verified that the
results obtained by this method do fulfill the sum rules with good accuracy
for all anisotropies and temperatures considered. The numerical data for
infinite temperature show that the logarithmic divergence of
$S^{zz}_{0,1}(\omega)$ for $\omega\to 0$ in the free fermion case becomes a
power-law $\sim\omega^{-\alpha}$ in the interacting case. In particular, I
found $\alpha\sim 0.3-0.4$ for $\Delta=1$. With decreasing temperature these
power-law divergencies become exponentially suppressed in $S^{zz}_0(\omega)$
as well as in $S^{zz}_1(\omega)$. For $\Delta=1$, I showed that the scale for
this exponential suppression is still set by $J$ as in the free fermion case.
The numerical data at low temperatures for $1/T_1^a$ (dominated by excitations
with wave-vector $q\sim\pi$) as well as the ones for $1/T_1^b$ (dominated by
excitations with wave-vector $q\sim 0$) are consistent with the field theory
predictions.

What does that mean for the NMR experiments on \Sr? First, concerning the
singular frequency dependence for $\omega\to 0$, there should be no difference
between measurements at the copper-, O(1)- or O(2)-site. Furthermore, a
singular frequency dependence should only show up when the temperature becomes
comparable to $J\approx 2000$ K. A diffusion-like contribution for $T\ll J$ as
has been suggested by Thurber {\it et al.}\cite{ThurberHunt} based on $^{17}$O
NMR measurements at the O(1)-site cannot be explained in a pure Heisenberg
model and with the hyperfine interaction being the only relevant relaxation
process.  However, the evidence presented in favor of such a contribution is
rather weak.  There is no reason to assume that the spin-lattice relaxation
rate in the limit of infinite magnetic field $h$ is given by the field theory
result where the effect of the magnetic field on the spin-spin correlations
has been ignored. In fact, the magnetic field can only be ignored if $T\gg
|h|$. Without this limiting value, however, the data in Fig.~3(d) of
Ref.~\onlinecite{ThurberHunt} are also consistent with having no frequency
dependence at all. If that is the case, the only part of the data which is not
consistent with the simple Luttinger model picture is the temperature
dependence of the relaxation rate at the O(1)-site, $1/(T_1^b\, T)\sim
\mbox{const}+ T$ compared to $1/(T_1^b\, T)\sim\mbox{const}$ expected from
field theory. As contributions from $q\sim\pi$ are strongly suppressed, this
next-leading temperature dependent term most likely has to do with corrections
to the simple $\delta$-peak (\ref{deltaFunc}) obtained from the Luttinger
model for the dynamical structure factor at small $q$. From recent studies at
zero temperature we know that a finite band curvature will broaden the
$\delta$-peak and lead to interesting singularities at the lower and upper
thresholds as well as to a high-frequency
tail.\cite{Glazman,Glazmannew,PereiraSirker} At finite temperatures, spectral
weight will also appear below the lower threshold. Further research, if these
corrections can indeed explain the measured temperature-dependence of
$1/(T_1^b\, T)$ is necessary. Finally, I want to remark that in the free
fermion case the next-leading term at low temperatures is of order $T^3$ (see
Eqs.~(\ref{f4},\ref{f5})) and not $T^2$. That suggests that if a
$T^2$-contribution exists, it should either have an amplitude which vanishes
for $\Delta=0$ or otherwise the exponent has to change as a function of
$\Delta$.  \acknowledgments I thank R.~G.~Pereira and I.~Affleck for useful
discussions.  This research has been supported by the German Research Council
({\it Deutsche Forschungsgemeinschaft}). The numerical calculations have been
performed using the Westgrid facilities.


\end{document}